\documentstyle[12pt,epsf]{article}
 1
\def\citenum#1{{\def\@cite##1##2{##1}\cite{#1}}}
\def\citea#1{\@cite{#1}{}}

\def\beq{\begin{equation}}
\def\eeq{\end{equation}}
\def\bea{\begin{eqnarray}}
\def\eea{\end{eqnarray}}

\def\undersum#1{\mathrel{\mathop{\sum}\limits_{#1}}}

\def\bbbz{{\mathchoice {\hbox{$\sf\textstyle Z\kern-0.4em Z$}}
{\hbox{$\sf\textstyle Z\kern-0.4em Z$}}
{\hbox{$\sf\scriptstyle Z\kern-0.3em Z$}}
{\hbox{$\sf\scriptscriptstyle Z\kern-0.2em Z$}}}}
\relax
\begin{document}
\begin{titlepage}
\noindent
 June 1994   \hfill  CBPF-NF-025/94 \\TAUP 2179 -94  \\ \\[9ex]
\begin{center}
{\Large \bf  The Role of Screening Corrections     }   \\[1.4ex]
{\Large \bf in High Energy Photoproduction
                        }   \\[11ex]

{\large E.\ Gotsman   *                         }    \\[1ex]
{\large E.M.\ Levin   ** $^{a)} $                          }    \\[1ex]
{\large U.\ Maor    *               }    \\[1.5ex]
{*  School of Physics and Astronomy  }      \\
{Raymond and Beverly Sackler Faculty of Exact Sciences} \\
{Tel Aviv University, Tel Aviv 69978}  \\ [1.5ex]
{** LAFEX, Centro Brasileiro de Pesquisas F\'\i sicas  (CNPq)}\\
{Rua Dr. Xavier Sigaud 150, 22290 - 180, Rio de Janiero, RJ, BRASIL}\\ [1.5ex]
\footnote{ a) On leave from Petersburg Nuclear Physics Institute,
Gatchina, St. Petersburg}
{\large \bf Abstract}
\end{center}

The role of screening corrections, calculated using the eikonal
model, is discussed in the context of soft photoproduction.
We present a comprehensive calculation considering the total,
elastic and diffractive cross sections jointly. We examine
the differences between our results and those obtained
from the supercritical Pomeron-Reggeon model with no
unitary corrections.
\end{titlepage}
\section {\bf  Introduction }
\par
The second generation of HERA results on total, elastic and diffractive
cross sections have recently become available \cite{1,2}. In this
note we explore the possibility of a comprehensive description of these
data which are considered to be soft processes. Both H1 and ZEUS
report that in the HERA energy domain ($ \sigma_{el}$ +
$ \sigma_{diff} $) is about 35-40\% of the total cross section.
This is similar  to the ratio observed in $ \bar{p}p$  scattering
in the ISR-Tevatron energy range, where the need for screening
corrections has been established \cite{3}. In the present
analysis we apply the methods  used in hadronic processes \cite{4,5}
to photon initiated processes, and attempt to formulate a
comprehensive description of total, elastic and diffractive
photoproduction reactions, in a Regge model with screening corrections.

 \par Our amplitude is
normalised so that
 \begin{equation}
   \frac{d \sigma}{dt} = \pi \vert f(s,t) \vert ^{2}
\end{equation}

 \begin{equation}
   \sigma_{tot} = 4 \pi Im f(s,0)
\end{equation}
 The scattering amplitude in {\it b}-space is defined as
 \begin{equation}
 a(s,b) = \frac{1}{2 \pi} \int d{\bf q}\;\; e^{-i{\bf q\cdot b}}
  f(s,t)
\end{equation}
where $ t= - q^{2} $ .    \\
 In this representation
 \begin{equation}
 \sigma_{tot} = 2 \int d{\bf b}\;\; Im a(s,b)
\end{equation}
 \begin{equation}
 \sigma_{el} = \int d{\bf b}\;\; \vert a(s,b) \vert^{2}
\end{equation}
\par
The introduction of screening rescattering corrections is greatly
simplified in the eikonal approximation where
  at high energy $ a(s,b) $ is assumed to be pure imaginary,
 and can be written in  the simple form
 \begin{equation}
a(s,b) = i ( 1 - e^{- \Omega(s,b)} )
\end{equation}
where the opacity  $ \; \Omega(s,b) $ is a real function.
As we shall utilize Regge parametrizations, analyticity and
 crossing symmetry are easily restored
  by substituting
$ s^{\alpha} \rightarrow s^{\alpha} e^{-i \pi \alpha /2}  $,
where $ \alpha $ denotes the exchanged Regge trajectory.

\par In  previous publications \cite{4,5} we have shown that
the eikonal approximation can be summed analytically for
 a Gaussian input
 \begin{equation}
 \Omega(s,b) =\nu(s) e^{-\frac{b^{2}}{R^{2}(s)}}
\end{equation}
Eq. (7) provides a good approximation for Regge type amplitudes,
where
\beq
Im f(s,t) = C e^{R^{2}_{0}t} (\frac{s}{s_{0}})^{\alpha(t) - 1}
 sin[\frac{ \pi \alpha(t)}{2}]
\eeq
with
$ \alpha(t) = \alpha(0) + \alpha^{ \prime} t $ .
On transforming to b-space we obtain

 \beq
\nu(s) = \frac{C R^{2}_{1}(s)}{2 \mid \beta \mid ^{2}}
(\frac{s}{s_{0}})^{\alpha(0) - 1}
 = \frac{\sigma_{0}}{2 \pi R^{2}(s)}(\frac{s}{s_{0}})^{\alpha(0) - 1}
\simeq \frac{\sigma_{tot}}{4 \pi B_{el}}
  \eeq
\beq
R^{2}(s) = \frac{4 \mid \beta \mid^{2}}{R^{2}_{1}(s)}
 \simeq 4 R^{2}_{1}(s)
\eeq
\beq
 R^{2}_{1}(s) =  R^{2}_{0} + \alpha^{\prime} ln (\frac{s}{s_{0}})
\eeq
\beq
\mid \beta \mid^{2} = R^{4}_{1} + \frac{ \pi ^{4} \alpha^{\prime \:2}}
{4}
\eeq
where $ \sigma_{0} = \sigma(s_{0})$ and $ B_{el} $
  denotes the elastic slope [$ B_{el} = \frac{1}{2} R^{2}(s)$ ].
With this input, we obtain in the eikonal approximation
 \beq
  \sigma_{tot} = 2 \pi R^{2}(s)[ ln \nu(s)
 + C - Ei(- \nu(s))]
\eeq
\beq
\sigma_{in} =
  \pi R^{2}(s) [ln 2 \nu (s) + C - Ei(- 2 \nu (s) ) ]
\eeq
\beq
\sigma_{el} = \sigma_{tot} - \sigma_{in}
\eeq
where
  $ Ei(x) = \int^{x}_{- \infty} \frac{e^{t}}{t} dt $  ,
and  C  = 0.5773 is the Euler constant.
In the following we take
$ \alpha_{P}(t) = 1 + \Delta + \alpha^{\prime} t $
and $ \alpha_{R} (t) = 0.5 +  t $.
\par The above parametrization also allows one to obtain
a closed expression for single diffraction dissociation , where
in the triple Regge limit with no screening corrections we have
\beq
\frac{M^{2}d \sigma_{sd}}{dM^{2}dt} =
\sigma_{0}^{2}(\frac{s}{M^{2}})^{2 \Delta + \alpha^{\prime}t}
[G_{PPP}(\frac{M^{2}}{s_{0}})^{\Delta}
+ G_{PPR}(\frac{M^{2}}{s_{0}})^{ - \frac{1}{2}}]
\eeq
With the introduction of screening corrections \cite{5}
we obtain
 \begin{eqnarray}
  \frac{M^{2}d \sigma_{sd}}{dM^{2}} =
  \frac{\sigma^{2}_{0}}{ 2 \pi {\bar R}^{2}_{1}(\frac{s}{M^{2}})}
         (\frac{s}{M^{2}})^{2 \Delta} \cdot
 [ G_{PPP}(\frac{M^{2}}{s_{0}})^{ \Delta}
  a_{1} \frac{1}{(2 \nu (s))^{a_{1}}} \gamma (a_{1},2 \nu (s))
   \nonumber    \\
 + G_{PPR} (\frac{M^{2}}{s_{0}})^{ - \frac{1}{2}}
  a_{2} \frac{1}{(2 \nu (s))^{a_{2}}} \gamma (a_{2},2 \nu (s)) ]
 \end{eqnarray}
where  $ G_{PPP} $ and $ G_{PPR} $ are the triple Regge
couplings corresponding to single diffraction dissociation,

 \beq
    {\bar R^{2}_{i}}(\frac{s}{M^{2}}) = 2 R^{2}_{0i} + r^{2}_{0i}
      + 4 \alpha^{ \prime} ln(\frac{s}{M^{2}})
 \eeq
   $ r_{0i} \leq 1 GeV^{-2} $ denotes the radius of the triple
 vertex and can safely  be neglected.
 \beq
a_{i} = \frac{2 R^{2}(s)}{ {\bar R}^{2}_{1}(\frac{s}{M^{2}}) +
     2 {\bar R}^{2}_{i}(\frac{M^{2}}{s_{0}})}
\eeq
The indices i =1,2 corresponds to P (Pomeron) and R (Reggeon)
exchanges.
 $ \gamma(a,2 \nu) $ denotes the incomplete Euler gamma function
$ \gamma(a,2 \nu) = \int^{2 \nu}_{0} z^{a - 1} e^{-z} dz $ .
\par
     The formalism presented above needs to be modified when applied
to photoproduction. To this end we make the following two assumptions:
\\
1) The photoproduction cross section can be  estimated using
the diagonal  vector dominance model (VDM) relation
\beq
  \sigma_{tot}(\gamma, p)\,=\,
\sigma_{VDM}(\gamma p)= \undersum{V = \rho,\omega,\phi}
\frac{4 \pi \alpha}{f^{2}_{V}} \sigma(Vp)
\eeq
where $ \frac{ 4 \pi \alpha}{f^{2}_{\rho}} \simeq \frac{1}{300} $,
and we assume the standard U-spin SU(3) relation
$ \rho : \omega : \phi $ = 9:1:2. \\
2) In addition, we assume the validity of the additive quark
model, where
\beq
\sigma_{tot}(\rho p) \simeq \sigma_{tot}(\omega p)
\simeq \frac{1}{2} [\sigma_{tot}(\pi^{+} p) +
\sigma_{tot}(\pi^{-} p) ]
\eeq
\beq
\sigma_{tot}(\phi p) \simeq \sigma_{tot}(K^{+} p)
+ \sigma_{tot}(K^{-} p) - \sigma_{tot}(\pi^{-} p)
\eeq
Using the Donnachie-Landshoff (DL) parametrization
of the  total cross sections \cite{6} we have \cite{7}
\beq
\sigma_{tot}(\rho p) \simeq \sigma_{tot}(\omega p)
\simeq 13.36 s^{\Delta} + 31.79 s^{- \eta}
\eeq
\beq
\sigma_{tot}(\phi p) \simeq 10.01 s^{\Delta} + 1.51 s^{- \eta}
\eeq
where $ \Delta =$ 0.0808 and $ \eta $ = 0.4525. With  these
numbers we deduce that in the HERA energy range the direct
$ \gamma-\rho $ coupling is responsible for 78\% of the
corresponding photoproduction cross section. At lower energies
the percentage of rho is slightly higher. In our calculations
we have used an overall average of $ C_{\rho} $ = 0.785,
$C_{\omega}$ = 0.090 and $C_{\phi}$ = 0.125.

\section {\bf Total cross sections}
\par  To facilitate a numerical calculation, we need to specify
our input to Eq. (9) and (10). To this end, and in accordance
with our basic assumption that the photoproduction processes
under consideration are soft, we utilize a DL type parametrization,
 where we take as our input
\beq
\sigma_{tot} = X (\frac{s}{s_{0}})^{\Delta}
  + Y (\frac{s}{s_{0}})^{- \eta}
\eeq
with  $ \Delta  $ = 0.0808 and $ \eta $ = 0.4525 \cite{6}. Our
choice X = 73.5 $ \mu$b and Y = 175 $ \mu$b ( with $ s_{0} $
= 1 $ GeV^{2} $ ) are bigger than those used by DL,
so as to compensate for the absorption initiated by the
eikonalization.
For Eq.(10)
we chose $ R^{2}_{0} $ = 4.6 $ GeV^{- 2} $ and
$ \alpha_{P}^{ \prime} $ = 0.25 $ GeV^{- 2} $. With this input
we are able to reproduce the low energy ( $ \sqrt{s} \leq $
15 GeV ) data well. Our results (denoted as GLM)
for the entire energy range
$ \sqrt{s} \geq $ 5 GeV are shown in Fig.1 together with
the relevant experimental points for $ \sigma_{tot}$ [1,2,8].
In general, the contribution of the Regge term
in the HERA energy range is negligible, so that these data
points actually fix the Pomeron term.
We did not attempt a " best fit" by fine tuning our parameters,
as considerable ambiguity still exists for the measurement
of $ \sigma_{tot} $ at HERA, where the reported experimental
errors are larger than 10\%, mostly due to  systematic uncertainties.

 A very important feature of our calculation
is that $ \sigma_{tot} $ in the energy range discussed,
behaves effectively like $ s^{0.066} $. This is lower than
our input value of $ \Delta $ = 0.0808 suggested by DL.
Our result is a direct consequence of the eikonalization summation,
where the input $ \Delta $ changes slowly with increasing
s, towards an asymptotic $ ln^{2}$s behaviour, as
implied by Eq. (13). Our value of $ \Delta_{eff} $ = 0.066
should be compared with ALLM \cite{9} who have
$ \Delta $ = 0.045 and Capella et al. \cite{10}
who find $ \Delta$ = 0.077. The exact value of
$ \Delta_{eff} $ in photoproduction processes is of
considerable interest. DL suggest a universal value of
$ \Delta $ = 0.0808. This is supported mainly by the
$ {\bar p}p$  data where $ \sigma_{tot} $ values are
available up to $ \sqrt{s}$ = 1800 GeV. As we  have noted
previously, the HERA data presently available
 is not sufficiently
 accurate to distinguish between the various choices
of $ \Delta_{eff} $. To be able to do this,
 the experimental error has to be reduced by about a factor
of three. This requires a radical reduction of the present
systematic error at HERA, which is not an easy task.

\section{\bf Elastic cross sections}
\par The photoproduction reaction $ \gamma$ + p $ \rightarrow $
V  + p ( where V = $\rho,\omega,\phi) $ and the total
cross section are related by
\beq
\frac{4 \pi \alpha}{f^{2}_{V}}  \sigma( \gamma p \rightarrow
 V  p) = ( 1 + \rho^{2}) \frac{[ C_{V} \sigma_{tot}(\gamma p)]
^{2}}{16 \pi B_{el}}
\eeq
 At HERA energies,
$ \: \rho^{2} \ll $ 1 and can safely be neglected. At
lower energies for the reaction
$ \gamma p \rightarrow \rho p $, we find that
$ \rho^{2} \simeq $ 0.1 for $ \sqrt{s} $ = 5 GeV,
 with increasing energy it reduces rapidly, so that at $ \sqrt{s}$
 = 20 GeV  $ \rho^{2} \simeq $ 0.01.
\par With our input Eqs.(20-24), and assuming the same
elastic slope $ B_{el}$ for $ \rho, \;\omega, \;$ and $ \phi $
photoproduction we obtain,
\beq
\sigma(\rho)\; : \sigma(\omega) \;: \sigma( \phi)
= 0.81 \; : 0.09 : 0.10
\eeq
Clearly, the best procedure  to analyze the elastic data
we are discussing, is to  make a combined analysis
of $ \sigma_{tot} $
,$ \sigma_{el}$ and $ B_{el} $ . This is not practical as
for $ \gamma p \rightarrow \rho p $ at low energies we have only one
  measurement \cite{11} of $ B_{el} $ = 10.6 $ \pm$  1.0 $ GeV^{- 2}$
 (at $ \sqrt{s} $ =14  GeV) ,
 which covers sufficiently low values of
$ \mid t \mid $. From the ZEUS data on $ \sigma_{tot} $
and $ \sigma_{el}$  quoted in Table I, we deduce that
$ B_{el} $ = 13.2 $ GeV^{- 2}$ at $ \sqrt{s}$ =180 GeV.
These values suggest
$ R^{2}_{0} $ = 4.0 $ GeV^{- 2} $. We have found
that a better overall reproduction of $ \sigma_{tot}$ and
$ \sigma_{el} $ is achieved with $ R^{2}_{0} $ = 4.6 $ GeV^{-2} $
, which is the value used throughout the present analysis.
\par The assumption that all vector mesons have a common slope
$ B_{el} $, is an
 oversimplification. From the analysis of purely hadronic
reactions we know that $ B_{el}$ becomes smaller with
increasing $ m^{2}_{V} $. For $ \gamma p \rightarrow \phi p $
we have a measured value \cite{12} of $ B_{el}\; = 6.0 \pm 0.3 \; GeV^{- 2} $
at $ \sqrt{s} $ = 2.5 -3.7  GeV . This corresponds to a choice
of $ R^{2}_{0} \simeq$ 2.3  $GeV^{- 2} $ and changes the ratio
given in Eq.(27) to
\beq
\sigma(\rho) \; : \sigma(\omega) \; : \sigma(\phi) =
0.79 \; : 0.08 \; : 0.12
\eeq
due to the different values of $ R^{2}_{0} $ chosen
for $ \rho, \omega $ and $ \phi$. We  note that in the DL model
$ B_{el}\; = \frac{1}{2} R^{2}(s) $. As a consequence of eikonalization
we expect that $ B^{GLM}_{el} > B^{DL}_{el}$. This difference which is
small at HERA energies, increases significantly with energy.

\par Our predictions for $ \sigma(\gamma p \rightarrow  \rho  p) $
together with the relevant data \cite{2,8} are illustrated in Fig.2.
We also show the predictions of the DL model taking
$ R^{2}_{0} $ = 4.6 $ GeV^{- 2}$. Once again, we observe
a systematic difference between the DL predictions and ours.
These differences are listed in Table I for $ \sqrt{s} $ = 180
GeV together with the reported ZEUS data \cite{2}. As in the
analysis of $ \sigma_{tot} $,
the error on the reported data point for
$ \sigma(\gamma p \rightarrow \rho p) $ = 18 $\pm$ 7 $\mu$b (
at $ \sqrt{s}$ =180 GeV ) is too large to allow one
to dicriminate between the two models.
\par Table I also contains  the predictions of the DL and GLM models
for $ \omega$ and $ \phi$ photoproduction. We have denoted
by DL1 and GLM1  the models evaluated with a common $ B_{el} $
, e.g. Eq.(27). By DL2 and GLM2 we denote the results where a different
value of the slope has been used for $ \phi$ photoproduction,
e.g. Eq.(28). The results for $ \gamma p \rightarrow \phi p $
should be compared with the parametrization suggested by DL,
where \cite{6}
\beq
\sigma(\gamma p \rightarrow \phi p ) = 0.275 \; s^{0.1616} \; (\mu b)
\eeq
The above parametrization leads to a value $ \sigma(\phi)$ = 1.5
$\mu$b at $ \sqrt{s}$ = 180 GeV. The problem is that the above fit
when compared to the DL parametrization for $ \sigma_{tot} $
implies that $ B_{el}$  = 18.0 $ GeV^{- 2} $  with no energy
dependence. This value is a factor of more than two larger
than the reported data point of Ref. 12.
\par In Table I we also show the results of our attempt to evaluate
$ \sigma( \gamma p \rightarrow \psi p) $ at $ \sqrt{s}$ = 180 GeV.
The data \cite{13} shows a rapid increase of $ \sigma(\psi) $
from threshold to a local plateau of 25 nb at $ \sqrt{s}$ = 15 GeV.
In our interpretation, this is the energy value where the Pomeron
contribution is larger than the Reggeon one, safely  enabling
us to extrapolate to $ \sqrt{s} $ = 180 GeV.  We use the same
notation as previously, i.e. DL1 and GLM1 correspond to a
common $ B_{el} $, which is not a realistic assumption due to the
relatively large mass of the $ \psi$.
For a more realistic estimate we use $ R^{2}_{0} $ = 0.5 $ GeV^{- 2} $
for DL2 and GLM2. Due to the small cross sections, the differnce
between the predictions for DL and GLM models are insignificant.
This is not suprising, as for small $ \Omega, $ (1 - $e^{- \Omega})
\simeq \; \Omega $.

 \section {\bf Diffractive cross sections}

\par There are three diffractive photoproduction channels to consider

\beq
 \gamma p \rightarrow X_{1} p
\eeq
\beq
\gamma p \rightarrow V X_{2}
\eeq
\beq
\gamma p \rightarrow X_{1} X_{2}
\eeq
The single diffractive channels given in Eqs.(30) and (31) can
be calculated using the triple Regge formalism, e.g. Eq. (16-17).
To perform such a calculation we need values for the parameters
$ G_{PPP}$ and $ G_{PPR} $, in addition to those parameters
fixed   from the analysis of the total and elastic cross sections.
Unfortunately, we do not have sufficient data to support such an
analysis. The only experimental data available \cite{11},
for reaction Eq.(30) is at  $ \sqrt{s}$ = 14 GeV, where
 a very simple triple critical Pomeron  ( $\Delta$ = 0) type
behaviour was assumed, and the PPR contribution was neglected.
With these assumptions one has
\beq
\frac{d \sigma_{sd}}{dt dx} = \frac{A}{1 - x}
\cdot e^{b(t + 0.05)}
\eeq
where the given fits for A and b enable us to evaluate
$ \sigma_{sd}( \gamma p \rightarrow X_{1} p)$ =
7.5 $\mu$b at this energy with the diffractive mass
2 $GeV^{2} \leq M^{2} \leq $ 0.05s.
\par If we assume a simple supercritical Pomeron model, such as the
DL model, then Eq.(16) implies that
\beq
\frac{\sigma_{sd}(s_{1})}{\sigma_{sd}(s_{2})}
=  ( \frac{s_{1}}{s_{2}})^{2 \Delta}
\frac{\bar{R^{2}_{1}}(\frac{s_{2}}{< M^{2}_{2}>})}
{\bar{R^{2}_{1}}(\frac{s_{1}}{< M^{2}_{1}> })}
\eeq
where $ <M^{2}_{i}> $ denotes the weighted average of the
$ M^{2} $ distribution.
\par The above expression for $ \sigma_{sd} $ grows much faster than
$ \sigma_{tot} $, so the need for screening corrections is
apparent. The energy scale at which these corrections
 become important is not known. Assuming Eq. (17),
we expect
\beq
\frac{\sigma_{sd}(s_{1})}{\sigma_{sd}(s_{2})}
=  ( \frac{s_{1}}{s_{2}})^{2 \Delta} \cdot
\frac{\bar{R^{2}_{1}}(\frac{s_{2}}{< M^{2}_{2}>})}
{\bar{R^{2}_{1}}(\frac{s_{1}}{< M^{2}_{1}> })} \cdot
\frac{[2 \nu(s_{2})]^{a(s_{2})}}
{[2 \nu(s_{1})]^{a(s_{1})}}
\frac{\gamma [a(s_{1}), 2\nu(s_{1})]}
{\gamma [a(s_{2}), 2\nu(s_{2})]}
\eeq
The energy dependence suggested by Eq.(35) is much milder than that
of the non-corrected  formula in Eq.(34). We note that
in the high energy limit $ B^{DL}_{sd} = \frac{1}{2}{\bar R^{2}_{1}} $
, while $ B^{GLM}_{sd} $ is slightly larger, however even at
HERA energies the difference is not significant.
\par To calculate the cross section for the reactions listed
in Eqs. (31) and (32), we assume the non-relativistic quark
model and approximate factorization, from which we derive
the equalities
\beq
\sigma_{sd}(\gamma p \rightarrow \rho X_{2})
= \frac{2}{3} \sigma_{sd}( \gamma p \rightarrow X_{1} p)
\eeq
\beq
\sigma_{sd}(\gamma p \rightarrow X_{1} X_{2}) =
\frac{\sigma_{sd}(\gamma p \rightarrow X_{1} p)
\sigma_{sd}(\gamma p \rightarrow \rho X_{2})}
{\sigma_{el}(\gamma p \rightarrow \rho p) }
\eeq
\par Two sets of results depending on the choice of $ R^{2}_{0} $
   for DL and for the unitarity corrected GLM
diffractive cross sections   at $ \sqrt{s}$ = 180 GeV are
summarized in Table I. Both the DL and our calculation
(GLM) yield a very reasonable diffractive cross section
of 29-35 $\mu$b. These results should be compared with the
experimental value
of 33 $\pm 8 \mu$b given by ZEUS.
\par We have also calculated the inclusive cross section for
$ \gamma + p \rightarrow \psi + X $. As in our previous calculations
we start with an average $ \sigma (\gamma + p \rightarrow \psi + X) $
= 18 nb at $ \sqrt{s} $ = 15 GeV \cite{14}, and obtain the results
given in Table I for $ \sqrt{s} $ = 180 GeV. Unfortunately, due to
the lack of  uniformity associated with the definition of diffractive
events, our results are somewhat uncertain.

\section {\bf Discussion}
\par The two models discussed in this  paper differ greatly in their
physical content. The DL model has a most attractive feature
in that all Pomeron exchange reactions have the same predicted energy
dependence, i.e. $ s^{0.0808} $ for  $ \sigma_{tot} $ and
$ s^{0.1616} $ modulated by a lns term for $ \sigma_{el} $ and
$ \sigma_{diff} $. One of the deficiencies  of the model is
that there is no hint
 at what energy scale this simple parametrization will fail to
 describe the data, in addition it has no provision for incorporating
unitarity corrections, which become important at higher energies.
Hence, the predictive power of the DL model at exceedingly high
energies is limited.
\par In contrast the GLM eikonal model is unitarized by construction,
and as such predicts that the effective energy dependence differs
in different energy domains, changing gradually from a powerlike
behaviour in energy to $ ln^{2}$s. In addition, the energy
dependence, which is universal in the DL model, evolves differently
for the different channels in the GLM eikonal model.
\par For this reason it is important to study photoproduction
processes and examine their energy dependence. At present
this is not a decisive test, as both the DL model with $ \Delta$
= 0.0808 and ALLM with $ \Delta$ = 0.045 are compatible
with the HERA data points. Our present treatment suggests
  an effective $ \Delta_{eff}$ = 0.066. We await improvement
of the data points measured at HERA, to discriminate between
the different models.
\par
In this paper we have studied photoproduction processes
initiated by a quasi real photon. The problem of how to
extrapolate these cross sections as a function of the photon's
virtuality $Q^{2}$,  has been discussed recently in several papers
[9,10,15]. A  conclusion common to  all of these
investigations is, that whereas $ \Delta$ is relatively small
( $\Delta \leq $ 0.1) for $ Q^{2} \leq $ 5 $GeV^{2}$,
it becomes considerably larger $ \Delta \simeq $ 0.35 for
$ Q^{2} > 10 \; GeV^{2} $. This has interesting consequences
for the case of high mass diffraction in real photoproduction
and DIS. Both H1 and ZEUS assume that
$ \frac{ d \sigma_{sd}}{d M^{2}} $ has a $ M^{- 2}$ dependence
on the diffractive mass. Actually, in the triple Regge
limit, with or without screening corrections, we expect the
dependence to be $ M^{- 2 \alpha_{P}(0)}$ ,
where $ \alpha_{P}(0) \; = 1 + \Delta$.  This was observed in
$ \bar{p}p$ interactions at the Tevatron \cite{16}. For real
photoproduction, where $Q^{2}$ = 0, this is an insignificant
  correction as $\Delta$ is small, however, for $Q^{2} > $ 10
$ GeV^{2}$ where $ \alpha_{P}$(0) = 1.35, we expect
 a dramatic change in the behaviour of
$ \frac{d \sigma_{sd}}{d M^{2}}$.

We would  like  to mention that  the GLM eikonal model
enables one to
 describe the matching between deep inelastic data with
 $\Delta\,\sim\, 0.35 $ and the photoproduction data with $\Delta \,\sim\,0.08$
incorporating sufficiently strong shadowing correction, as well
as non-Regge
 like power behaviour of the Pomeron contribution ( see Ref. [3] for the first
attempt in this direction). Since in QCD the Pomeron structure
is not a simple
Regge pole, we can incorporate the $Q^2$ behaviour both in the scale of the
shadowing correction, and in the effective power of
of $\nu$ on s. We postpone futher discussions of this
interesting question to further publications.
\par
{\bf Acknowledgements:}
We would like to thank all participants of Eilat conference for the
vigorous discussions on the subject. We are especially grateful to
J. Bjorken, A. Donnachie
and L.Frankfurt for sharing with us their view of the problems.
E.L. wishes to thank his colleagues at LAFEX/CBPF for their kind
hospitality and help, and is grateful to the CNPq for partial
financial support.
\newpage
\vglue 0.5cm
\vglue 0.4cm

\newpage
%
\section*{Table Captions}
\vglue 0.4cm
{\bf Table I.}  ZEUS data and the DL and GLM predictions
at $ \sqrt{s} $ = 180 GeV \\

\begin{center}
\begin{tabular}{|l|l|l|l|l|l|r}
   \hline
    & ZEUS data & DL1 & DL2 &GLM1 & GLM2 \\ \hline
$\sigma_{tot}\;( \mu$b)   &143 $\pm$ 4 $ \pm$ 17 & 157.9 & 157.9
& 149.8 & 149.8 \\ \hline \hline
$ \sigma(\gamma p \rightarrow \rho p) \; (\mu$b)  & 14.8 $\pm$ 5.7 &
17.1 & 17.1 & 13.7 & 13.7 \\ \hline
$\sigma(\gamma p \rightarrow \omega p) \; (\mu$b)  & & 1.9 & 1.9 & 1.5
 & 1.6 \\ \hline
$ \sigma(\gamma p \rightarrow \phi p ) \; (\mu$b)  & & 2.1 & 2.6 & 1.7
& 2.1 \\ \hline \hline
$\sigma(\gamma p \rightarrow V p )\; (\mu$b)  & 18.0 $\pm $7.0 & 21.1
& 21.6 & 16.9 & 17.4 \\ \hline \hline
$ \sigma(\gamma p \rightarrow X_{1} p) \; (\mu$b)  & &14.4  & 14.4
& 12.2 & 12.2 \\ \hline
$ \sigma(\gamma p \rightarrow V X_{2}) \; (\mu$b)  & & 12.0  & 12.2
& 10.2 & 10.3 \\ \hline
$\sigma(\gamma p \rightarrow X_{1} X_{2}) \; (\mu$b)  & & 8.1 & 8.2
& 7.3 & 7.4 \\ \hline \hline
$\sigma_{diff} \; (\mu$b)  & 33.0 $\pm$ 8.0 & 34.5 &34.8 & 29.7 &
29.9 \\ \hline \hline
$ \sigma(\gamma p \rightarrow \psi p) \;$ (nb)  & & 46.2 &31.5  &45.0
& 30.5 \\ \hline
$ \sigma(\gamma p \rightarrow \psi X) \; $(nb) & & 36.8 & 33.1 & 22.1
& 20.6 \\ \hline
\end{tabular}
\par
\end{center}
\newpage
\vglue 0.5cm
\section*{Figure Captions}
\vglue 0.4cm

{\bf Fig.1:}  Total photoproduction cross section as a function
of the $ \gamma$p center of mass energy. The solid line
is the prediction of the DL model, while the dashed line is
that of GLM.

{\bf Fig.2:} Cross section for $ \sigma( \gamma p \rightarrow \rho  p)$
as a function of the $ \gamma$p center of mass energy squared.
Curves as in Fig.1.
\end{document}